\documentclass[prb,showpacs,twocolumn]{revtex4}

\newcommand {\etal}{{\it et al.}}

\newcommand {\Hsf}{H_\textrm{SF}}
\newcommand {\Ppara}{\vec{P}_1}
\newcommand {\Pperp}{\vec{P}_2}
\newcommand {\BoldP}{\vec{P}}
\newcommand {\BoldH}{\vec{H}}

\newcommand{\SiSj}{\vec{S}_i \times \vec{S}_j}

\newcommand{\HAngle}{\theta_H}

\usepackage{graphicx}
\usepackage{bm}
\usepackage{pifont}
\usepackage{amsmath}

\begin{document}

\title{Cupric chloride CuCl$_2$ as an $S=1/2$ chain multiferroic}

\author{S. Seki$^1$, T. Kurumaji$^1$, S. Ishiwata$^{1,2}$, H. Matsui$^{3,4}$, H. Murakawa$^5$, Y. Tokunaga$^5$, Y. Kaneko$^5$, T.
Hasegawa$^4$ and Y. Tokura$^{1,2,5}$} 
\affiliation{$^1$ Department of Applied Physics, University of Tokyo, Tokyo 113-8656, Japan \\ $^2$ Cross-Correlated Materials Research Group (CMRG) and Correlated Electron Research Group (CERG), RIKEN Advanced Science Institute, Wako 351-0198, Japan \\  $^3$ Department of Advanced Materials Science, The University of Tokyo, Kashiwa, 277-8561, Japan  \\ $^4$ Photonics Research Institute (PRI), AIST, Tsukuba 305-8562, Japan \\ $^5$  Multiferroics Project, ERATO, Japan Science and Technology Agency (JST), Tokyo 113-8656}

\date{August 30, 2010}

\begin{abstract}

Magnetoelectric properties were investigated for an $S=1/2$ chain antiferromagnet CuCl$_2$, which turns out to be the first example of non-chalcogen based spiral-spin induced multiferroics. Upon the onset of helimagnetic order propagating along the $b$-axis under zero magnetic field ($H$), we found emergence of ferroelectric polarization along the $c$-axis. Application of $H$ along the $b$-axis leads to spin-flop transition coupled with drastic suppression of ferroelectricity, and rotation of $H$ around the $b$-axis induces the rotation of spin-spiral plane and associated polarization direction. These behaviors are explained well within the framework of the inverse Dzyaloshinskii-Moriya model, suggesting the robustness of this magnetoelectric coupling mechanism even under the strong quantum fluctuation.

\end{abstract}
\maketitle

\section{Introduction}

Multiferroics, materials with both magnetic and dielectric orders, have attracted revived interest.\cite{Review2} While the coupling between these orders are weak in general, recent discovery of spin-driven ferroelectricity in frustrated helimagnets has enabled unprecedentedly large magnetoelectric (ME) effects such as flop,\cite{Kimura, DyMnO3, MnWO4} reversal\cite{Rotation} or rotation\cite{EuY} of electric polarization ($P$) under applied magnetic field ($H$). Here, the key problem is the coupling mechanism between ferroelectricity and helimagnetism. The most successful scheme to explain such ME coupling is the inverse Dzyaloshinskii-Moriya (IDM) model,\cite{Katsura} which describes the local polarization $\vec{p}_{ij}$ produced between two magnetic moments $\vec{S}_i$ and $\vec{S}_j$ as:
\begin{equation}
\vec{p}_{ij} = A \vec{e}_{ij} \times (\SiSj),
\end{equation}
where $\vec{e}_{ij}$ is an unit vector connecting two magnetic sites, and $A$ is a coupling coefficient related to the spin-orbit interaction. Since the vector spin chirality $(\SiSj)$ is perpendicular to the spin-spiral plane, this model predicts that the $H$-induced tilt of spin-spiral plane leads to directional change of $P$. Ferroelectric (FE) and ME natures in several classical helimagnets like $R$MnO$_3$,\cite{Kimura, DyMnO3} Ni$_3$V$_2$O$_8$\cite{Ni3V2O8} and MnWO$_4$\cite{MnWO4} are successfully explained by this IDM scheme.

\begin{figure}
\begin{center}
\includegraphics*[width=8.5cm]{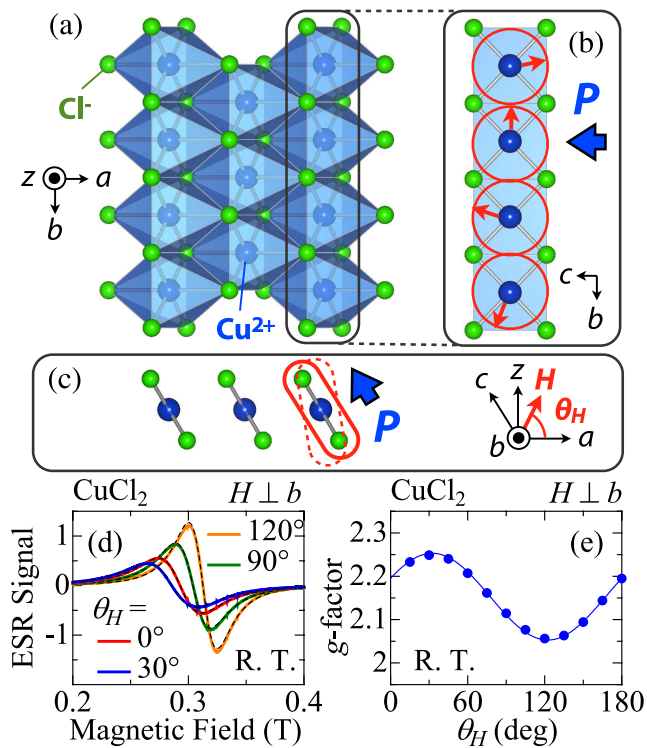}
\caption{(color online). (a)-(c) Crystal structure of CuCl$_2$, and $P$-direction observed at the ground state. The $bc$-cycloidal spin order suggested by Banks {\etal}\cite{CuCl2_Banks} is illustrated in (b), and also in (c) with solid rounded square representing spin-spiral plane. Dashed rounded square indicates the possible tilting of spin-spiral plane as revealed in this study(see text). (d) ESR signal taken at room temperature under various directions of $H$ confined within the $ac$-plane. $\theta_H$ is defined as the angle between the $a$-axis and $H$-direction. Each dashed line represents a fitted curve with a single Lorentzian resonance. (e) Angle dependence of $g$-factor.}
\label{fig1}
\end{center}
\end{figure}

In contrast, still in controversy is the ME coupling mechanism in $S=1/2$ chain magnets, where strong quantum fluctuation is believed to have some profound effects on their ME response.\cite{QHelicity, Furukawa} Typical examples are LiCu$_2$O$_2$ and LiCuVO$_4$, both of which are characterized by edge-shared CuO$_2$ chains. While their simple crystal structures and reported helimagnetism are seemingly typical of the IDM scheme, the experimentally observed $P$-direction or ME response for applied $H$ appears to contradict with its prediction. For example, LiCu$_2$O$_2$ hosts $P \parallel c$ in the helimagnetic ground state,\cite{LiCu2O2_First} but its proposed magnetic structures are contradictory among several experiments.\cite{LiCu2O2_Masuda, LiCu2O2_PolarizedNeutron, LiCu2O2_SoftXRay, LiCu2O2_Yasui} Even if we simply assume $bc$-spin spiral consistent with the IDM model, $P$-behaviors under applied $H$ contradict with its naive prediction.\cite{LiCu2O2_First} In case of LiCuVO$_4$, $P \parallel a$ is found in $ab$-spiral spin phase at 0 T, consistent with the IDM model.\cite{LiCuVO4_First, LiCuVO4_NeutronOne} However, the observation of $P \parallel c$, assigned to the $bc$-spiral spin phase,\cite{LiCuVO4_Loidl} was not reproduced by another group.\cite{LiCuVO4_NeutronYasui} 
Since both compounds frequently contains Li-Cu intersubstitution due to their close ionic radii, Moskvin {\etal} have claimed that the observed FE and ME natures stem purely from exchange striction and crystallographic defects, not from the spin-orbit interaction (or the IDM scheme).\cite{KNBCancel, NonstOne, NonstTwo} Furthermore, some recent theoretical study predicted that quantum fluctuation may largely reduce the effective magnitude of $P$ induced via the spin-orbit interaction.\cite{Furukawa} 
To testify the validity of the IDM model in quantum chain magnets, it is crucial to check the ME response in other $S=1/2$ compounds with similar edge-shared chain structures.

Our target compound, anhydrous cupric chloride CuCl$_2$ crystalizes into distorted CdI$_2$ form with monoclinic $C2/m$ space group and $\beta =122 ^\circ$.\cite{CuCl2_Structure} While original CdI$_2$ structure consists of the stacking of triangular lattices along the $z$-axis,\cite{CuCl2_Comment} they are extended along the $a$-axis due to Jahn-Teller active Cu$^{2+}$ ions(Fig. \ref{fig1}(a)). As a result, CuCl$_2$ can be regarded as the aggregate of edge-shared chains running along the $b$-axis, with CuCl$_4$ square plaquettes lying in the $bc$-plane (Fig. \ref{fig1}(b)). Magnetism is dominated by the intra-chain coupling between Cu$^{2+}$ ($S = 1/2$) ions, and competition between ferromagnetic nearest-neighbor interaction and antiferromagnetic next-nearest-neighbor interaction stabilizes the helimagnetic ground state below 24 K.\cite{CuCl2_TN, CuCl2_Water, CuCl2_Banks} Recent powder neutron scattering study suggested the cycloidal magnetic order propagating along the $b$-axis, with spin spiral confined in the $bc$-plane (Fig. \ref{fig1}(b)) and propagation vector $q\sim(1, 0.226, 0.5)$.\cite{CuCl2_Banks} While no dielectric measurements have been reported, the latest calculation based on density functional theory (DFT) predicts emergence of ferroelectricity along the $c$-axis.\cite{CuCl2_Banks} In this paper, we report the experimental discovery of FE and ME natures in CuCl$_2$, and prove that the IDM mechanism is still robust even under the strong quantum fluctuation. CuCl$_2$ is also the first example of non-chalcogen based spiral-spin induced multiferroics, which promises further discovery of unique ME function in many $MX_2$-type compounds and other forms of halide compounds.

\section{Experiment}

Single crystals of CuCl$_2$ were grown by a Bridgman method. They were cleaved along planes perpendicular to the $z$-axis, and cut into a rectangular shape with additional faces perpendicular to the $a$- or $b$-axis. Silver paste was painted on end surfaces as electrodes. Due to its moisture sensitivity, the specimen was handled in an Ar-filled glove box. To deduce $P$, we measured the polarization current with constant rates of temperature($T$)-sweep ($5\sim 20 $ K/min), $H$-sweep (100 Oe/sec), or $H$-rotation (2$^\circ$/sec), and integrated it with time. To enlarge the population of specific $P$-domain, the poling electric field ($E=150 \sim 400$ kV/m) was applied in the cooling process and removed just prior to the measurements of polarization current. Dielectric constant $\epsilon$ was measured at 1 MHz using $LCR$-meter. Magnetization $M$ was measured with a SQUID magnetometer. ESR signal was measured by JEOL JES-FA200 at X-band frequency ($\sim 9.0$ GHz).

\section{Results and discussions}

\begin{figure}
\begin{center}
\includegraphics*[width=8.5cm]{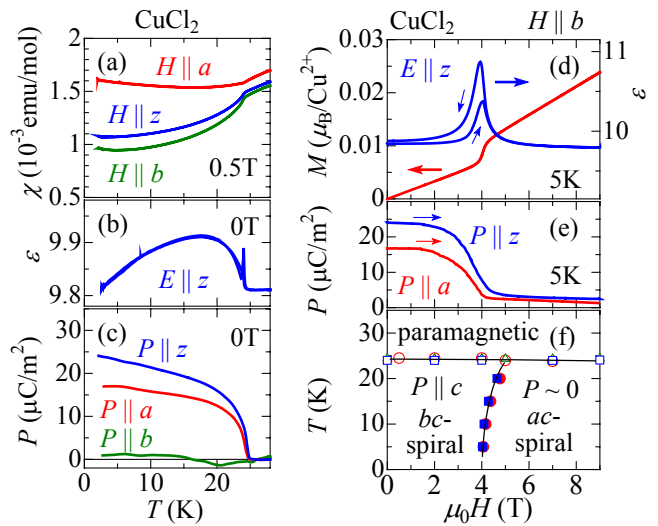}
\caption{(color online). Temperature dependence of (a) magnetic susceptibility $\chi$, (b) dielectric constant $\epsilon$, and (c) electric polarization $P$. In (d) and (e), $H$-dependences of magnetization $M$, $\epsilon$, and $P$ under $H \parallel b$ are indicated. Large and small arrows denote corresponding ordinate scale of physical quantity and direction of field scans,  respectively. (f) $H$-$T$ phase diagram for $H \parallel b$. Circles, squares, triangles are the data points obtained from $M$, $\epsilon$ and $P$ profiles, respectively. Open (closed) symbols are taken from $T$- ($H$-) increasing runs.}
\label{fig2}
\end{center}
\end{figure}

As suggested in Ref. 27, the $ac$-twin domains are expected to readily occur. To check this possibility, we first performed ESR measurements under various directions of $H$ confined within the $ac$-plane(Fig. \ref{fig1}(d)). Hereafter, we define $\HAngle$ as the angle between the $a$-axis and $H$-direction. Each observed profile can be fitted well with a single Lorentzian resonance for all $\HAngle$, indicating our crystal grown by the Bridgman method has no crystallographic twinning. The deduced $g$-factor shows sinusoidal $\theta_H$-dependence(Fig. \ref{fig1}(e)), whose maximum and minimum values agree well with those previously reported for a twinned crystal.\cite{CuCl2_Banks}

\begin{figure}
\begin{center}
\includegraphics*[width=6cm]{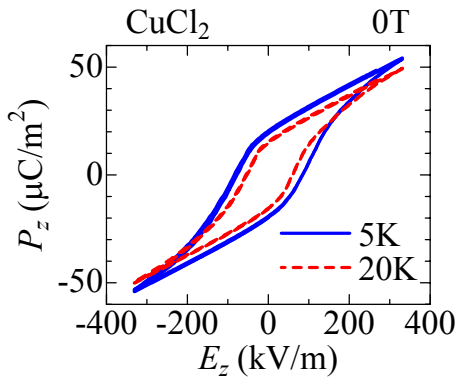}
\caption{(color online). $P$-$E$ hysteresis loop for CuCl$_2$ measured with $E \parallel z$. Electric field was swept at the rate of 13kV/m$\cdot$sec. }
\label{fig3}
\end{center}
\end{figure}

Next, we measured $T$-dependence of magnetic susceptibility $\chi$, $\epsilon$, and $P$ (Figs. \ref{fig2}(a)-(c)). $\chi$ suddenly drops at $T_\textrm{N} \sim 24$ K, which signals the transition into a spiral magnetic phase. Simultaneously, $z$-component of $\epsilon$ ($\epsilon_z$) shows a sharp anomaly and $a$- and $z$-components of $P$ ($P_a$ and $P_z$) begin to develop. The direction of $P_z$ can be reversed with reversal of applied $E$, producing the typical $P$-$E$ hysteresis curve (Fig. \ref{fig3}), while no $P_b$ component could be confirmed. These results imply strong correlation between helimagnetic and FE orders in CuCl$_2$. Based on the $bc$-plane helimagnetic structure suggested in Ref. 27, the IDM model as well as the DFT calculation\cite{CuCl2_Banks} predicts  $|P_a / P_z| \sim 0.64$ (i.e. $P \parallel c$). This roughly agrees with the observed $|P_a / P_z| \sim 0.70$.

Figures \ref{fig2}(d) and (e) indicate $H$-dependence of $M$, $\epsilon$, and $P$ for $H \parallel b$. At 4 T, $M$-profile shows a clear step as already reported.\cite{CuCl2_Banks, CuCl2_MH} Concomitantly, $\epsilon_z$ shows a sharp peak and both $P_a$ and $P_z$ are drastically suppressed. Since antiferromagnetic spins favor to lie within a plane perpendicular to $H$, this transition should correspond to a spin-flop into the $ac$-spiral spin state. The $ac$-spiral spin structure belongs to a magnetic form called proper screw, where spin-spiral plane is perpendicular to the modulation vector along the $b$-axis. The IDM model predicts $P = 0$ for this type of spin order due to the relationship $(\SiSj) \parallel \vec{e}_{ij} \parallel b$, which is consistent with the observed suppression of $P$. Figure \ref{fig2}(f) summarizes the obtained $H$-$T$ phase diagram for $H \parallel b$. The boundary of  the FE phase always coincides with that for magnetic phases, which proves the interplay between FE and helimagnetic natures. 

\begin{figure}
\begin{center}
\includegraphics*[width=8.5cm]{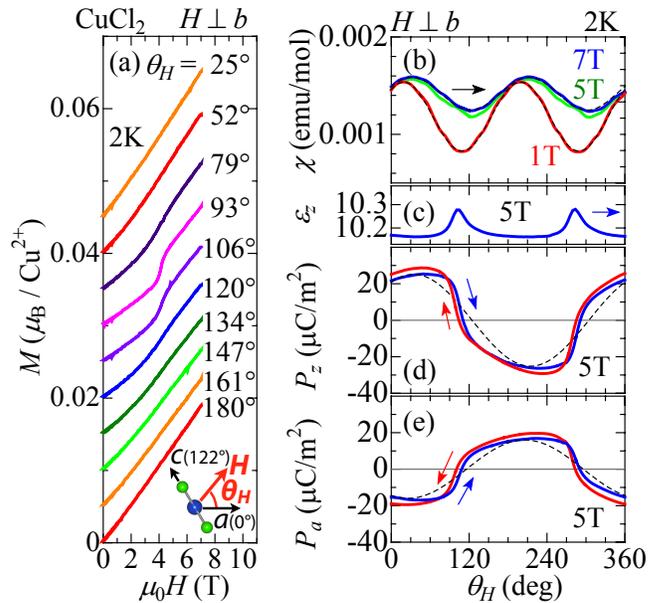}
\caption{(color online). (a) $H$-dependence of $M$ taken under various directions of $H$ around the $b$-axis. The base lines of data are arbitrarily shifted. (b) Angle-dependence of $\chi$ measured under $H \perp b$. Dashed lines indicate the fits with sinusoidal function. (c)-(e) Corresponding development of $z$-component of $\epsilon$ ($\epsilon_z$) as well as $z$- and $a$-components of $P$ ($P_z$ and $P_a$). Before measurements of $P$, the specimen was cooled at $\theta_H = 0$ with poling $E$ applied along the $z$-axis. Dashed lines indicate the behavior expected from Eq. (2). Arrows indicate the direction of $H$-rotation.}
\label{fig4}
\end{center}
\end{figure}

We further investigated the properties under $H \perp b$. Here, we adopt the same definition of $\theta_H$ as used for ESR measurements. Figure \ref{fig4}(a) indicates $H$-dependence of $M$ measured at various $\theta_H$. While no magnetic transition has been reported for $H \perp b$,\cite{CuCl2_MH} we found a clear signature of spin-flop at $\Hsf \sim 4$ T most pronounced around $\theta_H = 100^\circ$. $\theta_H$-dependence of $\chi (=M/H)$ was also measured (Fig. \ref{fig4}(b)), and $\chi$ sinusoidally changes with minimum at $\theta_H \sim 100^\circ$ below $\Hsf$. In general, the sharpest transition of spin-flop as well as the minimum value of $\chi$ should be observed when $H$ is applied parallel to the magnetic easy-plane. These results imply the magnetic easy-plane, i.e. spin-spiral plane at the ground state, is tilted from the originally suggested $bc$-plane toward the $bz$-plane by about $20^\circ$(Fig. \ref{fig1}(c)). Above $\Hsf$, $\chi$ still modulates sinusoidally but with different $\chi$-minimum position at $\theta_H \sim 122^\circ$ (i.e. $H \parallel c$). With $H > \Hsf$, the gain of Zeeman energy exceeds the energy loss due to magnetic anisotropy, and continuous rotation of spin-spiral plane is expected. In this case, $\theta_H$-dependence of $\chi$ rather reflects the anisotropy of $g$-value,\cite{Nagamiya} whose minimum is also confirmed to appear at $H \parallel c$ (Fig. \ref{fig1}(e)).

\begin{figure}
\begin{center}
\includegraphics*[width=8.5cm]{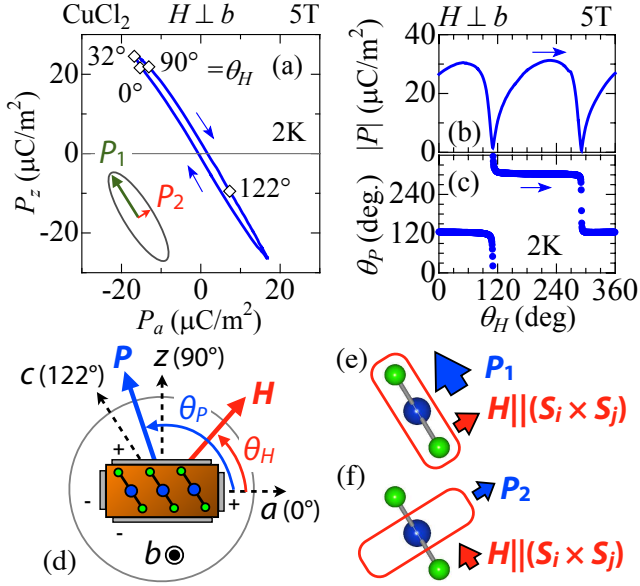}
\caption{(color online). (a) Trace of $P_a$ and $P_z$ under $H$ rotating around the $b$-axis. (b),(c) Magnitude and direction of $P$ as a function of $H$-angle. Arrows indicate the direction of $H$-rotation. The data are taken from Figs. \ref{fig4} (d) and (e), and the setup for measurements is shown in (d). $\theta_P$ ($\theta_H$) is defined as the angle between $P$- ($H$-) direction and the $a$-axis. (e),(f) The expected relationship between $P$, $H$ and spin-spiral plane (depicted as rounded square).  } 
\label{fig5}
\end{center}
\end{figure}

To investigate the behavior of $P$ under $H$ rotating around the $b$-axis, we simultaneously measured $P_z$ and $P_a$ using two pairs of electrodes. Thus, both $P$ and $H$ can be expressed as vectors within the $ac$-plane. We also define $\theta_P$ as the angle between the $a$-axis and observed $P$-direction (Fig. \ref{fig5}(d)). Figures \ref{fig4}(d) and (e) indicate $\theta_H$-dependences of $P_z$ and $P_a$ measured at 5 T. When $H$ is rotated by 180$^\circ$, $P$-direction is always found to be reversed. To see the behavior of $P$ more straightforwardly, the trace of $P$ is plotted in the $P_a$-$P_z$ plane(Fig. \ref{fig5}(a)). It forms a shape like elongated ellipse.  In Figs. \ref{fig5}(b) and (c), $\theta_H$-dependences of $|P|$ (magnitude of $\BoldP$) and $\theta_P$ are indicated. $\theta_P$ takes almost constant value around $\theta_P = 120^\circ$ or $300^\circ$, suggesting the major axis of observed $P$-ellipse is pointing at the $c$-axis.  If we assume that $H$ is always perpendicular to the spin-spiral plane, i.e. $\BoldH \parallel (\SiSj)$, the IDM model predicts $\BoldP = \Ppara \parallel c$ for $\BoldH \perp c$ (Fig. \ref{fig5}(e)) and $\BoldP = \Pperp \perp c$ for $\BoldH \parallel c$ (Fig. \ref{fig5}(f)). For general $\theta_H$, $\BoldP$ is given as 
\begin{equation}
\BoldP = \Ppara \sin (122^\circ - \theta_H) + \Pperp \cos (122^\circ - \theta_H),
\end{equation}
which forms an ellipse-shaped trace with $\Ppara$ and $\Pperp$ as the major and minor axes, respectively. From the $|P|$-profile, we deduced $| \Ppara | \sim$ 31 $\mu$C/m$^2$ and $| \Pperp | \sim$ 2 $\mu$C/m$^2$.

In Figs. \ref{fig4}(d) and (e), the $P$-behavior expected from Eq. (2) is plotted as dashed lines. While the calculated $P$-profile roughly agrees with the observed one, small gap still exists between them. This deviation reverses its sign at $\theta_H \sim 100^\circ$, where $H$ becomes parallel to the magnetic easy-plane. Correspondingly, $\epsilon$ also shows small anomaly at $\theta_H \sim 100^\circ$ (Fig. \ref{fig4}(c)). These behaviors can be well explained by assuming that the spin-spiral plane is tilted from the originally expected $\BoldH \parallel (\SiSj)$ position toward the magnetic easy-plane. A similar effect of magnetic-anisotropy drag on $P$ has also been observed in the $H$-rotating experiment on Eu$_{1-x}$Y$_x$MnO$_3$.\cite{EuY}

Thus, we conclude that the IDM scheme can well reproduce the observed FE and ME natures, even for CuCl$_2$ with $S=1/2$ quantum spin chains. Note that $P \parallel c$ relationship observed at 0 T can be justified even with a slight revision of the originally suggested $bc$-spiral spin structure, since deduced ratio $|\Ppara|/|\Pperp| \sim 15$ is quite large. Notably, when $H$ and spin-spiral plane is rotated counter-clockwise, $P$ is found to rotate clockwise (Fig. \ref{fig5}(a)). This is in contrast with the case for Eu$_{1-x}$Y$_x$MnO$_3$,\cite{EuY} where both $H$ and $P$ rotate in the same direction. The observed manner of $P$-rotation and large $|\Ppara|/|\Pperp|$ ratio are in accord with the recent DFT calculation for edge-shared CuO$_2$ chain compounds,\cite{CuO2_DFT} and these features would reflect the anisotropy and sign of coupling coefficient $A$ in Eq. (1).


CuCl$_2$ is also the first example of non-chalcogen based spiral-spin induced multiferroics. While observed $|\Ppara| \sim$ 31 $\mu$C/m$^2$ is somewhat smaller than the calculation ($|\Ppara| \sim$ 84 $\mu$C/m$^2$),\cite{CuCl2_Banks} it is comparable with the case for other helimagnetic oxides ($2000 \sim$ 5 $\mu$C/m$^2$). Interestingly, a recent theoretical study suggested the choice of anion may largely affect the value of induced $P$ through the different strength of metal-ligand hybridization and spin-orbit coupling.\cite{Jia1} This means that the ME response can be enhanced if we choose appropriate anion as the ligand. Until now, the study of ferroelectric helimagnets is almost limited to oxides, partly because their isostructural chalcogen relatives with larger anions (i.e. sulfides or selenides) are often electrically too leaky to perform dielectric measurements.\cite{CuCrO2_Seki, AgCrS2, CdCr2S4} Since halogens have larger electronegativity than chalcogens, halides are better insulating and enable the investigation of ME properties for a wider variety of anions. For example, most of $MX_2$-type halides with $X =$ Cl, Br, and I consist of stacking of undistorted triangular lattices, which realizes various types of spiral spin orders.\cite{VCl2, MnI2, CoI2} The systematic investigation of FE properties of whole $MX_2$ system will offer a good opportunity to clarify the anion dependence of magnetically-induced ferroelectricity, which may contribute to a general strategy to obtain larger ME responses.

\section{Conclusions}

In summary, we have experimentally revealed magnetically-driven ferroelectricity in an $S=1/2$ chain helimagnet CuCl$_2$. Observed $P$-behaviors under applied $H$ can be reproduced well within the framework of the inverse Dzyaloshinskii-Moriya model, suggesting the robustness of this ME coupling mechanism even under the effect of strong quantum fluctuation. CuCl$_2$ is the first example of non-chalcogen based spiral-spin induced multiferroics, which promises further discovery of unique magnetoelectric function in many $MX_2$-type compounds and other forms of halide compounds.

\begin{acknowledgments}

The authors thank T. Arima, N. Nagaosa, Y. Kohara, and Y. Taguchi for enlightening discussions. This work was partly supported by MEXT of Japan and JSPS through Grants-In-Aid for Scientific Research (Grant No. 20340086, 2010458) and Funding Program for World-Leading Innovative R\&D on Science and Technology (FIRST Program).

\end{acknowledgments}


\begin{thebibliography}{100}
\bibitem{CuCl2_Comment} For simplicity, hereafter we define the $z$-axis as the direction perpendicular to both $a$- and $b$-axes(Fig. \ref{fig1}(c)).
\end{thebibliography}
\end{document}